\documentclass[a4paper]{article}

\usepackage{INTERSPEECH2019}
\usepackage{subfig}
\usepackage{multirow} 
\usepackage{array}
\usepackage{xcolor}
\usepackage{hyperref}
\newcolumntype{P}[1]{>{\centering\arraybackslash}p{#1}}
\newcolumntype{M}[1]{>{\centering\arraybackslash}m{#1}}

\title{Does the Lombard Effect Improve Emotional Communication in Noise? \\ -- Analysis of Emotional Speech Acted in Noise --}
\name{Yi Zhao$^1$, Atsushi Ando$^2$, Shinji Takaki$^1$, Junichi Yamagishi$^1$, Satoshi Kobashikawa$^2$}
\address{
  $^1$National Institute of Informatics, Japan
  $^2$NTT Media Intelligence Laboratories, Japan}
\email{zhaoyi@nii.ac.jp, ando.atsushi@lab.ntt.co.jp, takaki@nii.ac.jp,\\ jyamagis@nii.ac.jp, kobashikawa.satoshi@lab.ntt.co.jp}

\begin{document}

\maketitle
\ninept
\begin{abstract}
Speakers usually adjust their way of talking in noisy environments involuntarily for effective communication. This adaptation is known as the Lombard effect.
Although speech accompanying the Lombard effect can improve the intelligibility of a speaker's voice, the changes in acoustic features (e.g.\ fundamental frequency, speech intensity, and spectral tilt) caused by the Lombard effect may also affect the listener's judgment of emotional content.
To the best of our knowledge, there is no published study on the influence of the Lombard effect in emotional speech.
Therefore, we recorded parallel emotional speech waveforms uttered by 12 speakers under both quiet and noisy conditions in a professional recording studio in order to explore how the Lombard effect interacts with emotional speech.
By analyzing confusion matrices and acoustic features, we aim to answer the following questions:
1) Can speakers express their emotions correctly even under adverse conditions?
2) Can listeners recognize the emotion contained in speech signals even under noise?
3) How does emotional speech uttered in noise differ from emotional speech uttered in quiet conditions in terms of acoustic characteristic?
\end{abstract}

\section{Introduction}
Humans adapt their speech to the physical environment. 
To avoid degradation of speech intelligibility in a noisy environment, speakers usually adjust the way they talk for effective communication. This involuntary adaptation is commonly referred to as the Lombard effect~\cite{lane1971lombard}. Lombard speech, which refers to speech produced in the presence of noise, is known to be more intelligent than `normal' speech when presented in equivalent amounts of noise~\cite{dreher1957effects,summers1988effects,lu2008speech,pittman2001recognition,garnier2014speaking}.
This is the natural result of the feedback system between vocal production and auditory perception that enables correction of speech performance~\cite{zollinger2011lombard}. 

The regular changes between normal and Lombard speech include not only loudness but also other acoustic features, such as
prolonging the duration of their speech~\cite{villegas2012role}, shifting the timing of vocalizations, increasing the pitch of vocalisations~\cite{patel2008influence,lu2009contribution}, shifting in formant center frequencies for F1 (mainly) and F2~\cite{kirchhuebel2010effects}, and shifting in energy from low-frequency bands to middle or high bands~\cite{garnier2006acoustic}. Is has also been demonstrated that spectral tilt decreases, implying an increase in high-frequency components under the Lombard effect~\cite{lu2009contribution}. 
Also, Lombard changes are greater in adults than in children, and in spontaneous speech than in reading tasks~\cite{amazi1982lombard}. 

In addition to its effect on psychophysiology, hearing tests, and studies on audio-vocal integration~\cite{zollinger2011lombard},
the Lombard effect has significant impacts  in applications of speech-related technology, such as  speaker recognition~\cite{hansen2009analysis}, noisy speech recognition~\cite{hansen1990lombard, chi1996lombard}, Lombard speech synthesis~\cite{raitio2011analysis,raitio2014synthesis}, and so on. 
It is also applicable to the study of vocal disorders and speech production, and has even been used as a therapeutic tool to improve speech intelligibility in Parkinson's disease patients \cite{zollinger2011lombard}.
In architectural acoustics and design, studies on the Lombard effect have been utilized to reduce unwanted noise and improve intelligibility of speech indoors~\cite{kleijn2015optimizing}. The Lombard effect is relevant to the study of phonetics and linguistics \cite{zollinger2011lombard}, too.


Although the Lombard effect has been widely studied, we still know far too little about the relationship between its mechanisms and human behavior, particularly when it comes to emotional speech.
Emotions color the language and act as a necessary ingredient for natural human communication and interaction. 
With the development of artificial intelligence, machines are expected to not only understand  human speech but also to be capable of capturing human emotions and generating emotional synthetic speech just like a real human being.
There is a wide range of studies on emotional speech in areas such as speech recognition and speech synthesis. 
However, to the best of our knowledge, there's no published study to quantify any influence between the Lombard effect and emotion expression in human speech.

In this paper, we investigate how the Lombard effect affects emotional speech from both speaker and listener perspectives on the basis of confusion matrix and acoustic analysis. 
For this purpose, we recorded a Japanese emotional database that contains four  emotions (happy, sad, angry, neutral) acted under both quiet and noisy conditions\footnote{Speech samples are available at:\\ https://nii-yamagishilab.github.io/EmotionaLombardSpeech}.
The context of this database is balanced in terms of both emotions and environment.
While recording emotional speech uttered by speakers, several listeners were asked to identify the  emotion contained in the actor's speech at once, and their feedback was given to the actor to improve the actor's expressions.

After recording the speech database, we first analyzed the accuracy of the actor's performance and listeners' judgment on the basis of confusion matrices~\cite{ting2017confusion} and Frobenius distance~\cite{amendola2015model,laurent2012forecasting}. 
Moreover, we also statistically analyzed the acoustic changes of speech produced in quiet and noisy conditions. 
The experimental results reveal many interesting phenomena.


In Section 2 of this paper, we briefly introduce the recording procedure of emotional speech acted in quiet/noisy environments. 
In Section 3, we try to answer the first two questions mentioned in the abstract from analysis results. 
We analyze the confusion matrix of acted/perceived emotions, and calculate Frobenius distance according to confusion matrices from various perspectives.  
In Section 4, we show the distributions of acoustic features for different emotions in either quiet or noisy environments using statistical methods.
In Section 5, we measure the speech intelligibility of recorded emotional speech by using a standard objective measure.
We conclude in Section 6 with a brief summary and mention of future work.
\section{Recording Procedure of Emotional Speech Acted in Quiet/Noisy Environment}
\subsection{Participants}
Twelve trained Japanese native speakers (six males and six females) aged 20\textendash{}40 years took part in the recording. 
None of them had any basic knowledge about the Lombard effect. None reported any speech or hearing difficulties.
These speakers are all professionally educated voice actors, six of them certified as 'high-level' and the other six as 'low-level'.
Three different listeners were assigned to identify the perceived emotion of each utterance for each speaker during the recordings.
In total, 36 listeners (18 males and 18 females) aged 20\textendash{}60 years old (ten in their 20s, ten in their 30s, eight in their 40s, eight in their 50s) participated in the judgments.

\subsection{Task}
All recordings in both quiet and noisy environment conditions use the same set of ten parallel sentences with different contexts, and each sentence is asked to be uttered in four different emotions: happy, sad, angry, and neutral. The order of sentences and emotions are randomized during the recording.
For each speaker, we have at least 80 successfully performed and correctly pronounced utterances (40 in the quiet environment and 40 in the noisy one).
The speech material comprises at least 960 sentences (4 emotions $\times$ 12 actors $\times$ 10 sentences $\times$ 2 conditions).

\begin{figure}[t]
\centering
\includegraphics[width=8cm]{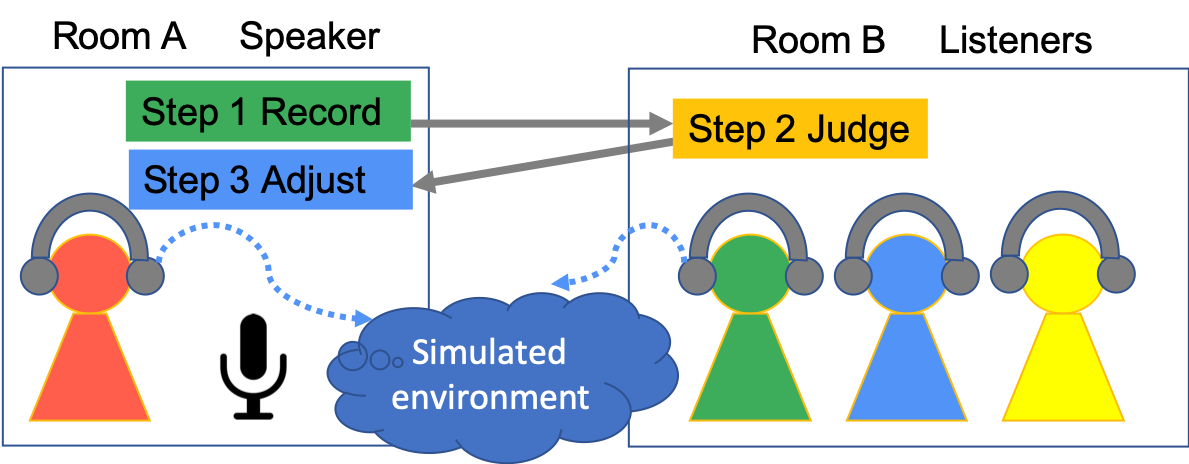}
\vspace{-0.5cm}
\caption{Recording procedure}\label{fig:recording}
\vspace{-0.4cm}
\end{figure}

The recording procedure is shown in Fig.~\ref{fig:recording}. To account for the influence of emotional speech with the Lombard effect, speakers and listeners were recorded while playing a collaborative game in pairs.
During recording, the speaker was first asked to read one utterance with the assumed emotion, the pronunciation of the recorded speech was then quickly verified, and the correctly pronounced speech was sent to the listeners' headsets. The listeners were asked to write down the perceived emotion and they could select `unknown' if they were indecisive. Their judgments were sent back to the speaker to help him/her adjust the performance.
Speakers and listeners were arranged in different rooms to make sure they could not hear each other. 
Also, listeners could not see each other during the experiment.

During the recording of emotional speech in a noisy environment, the speakers were asked to listen to noise played thorough their headphones while they read out the sentences. Listeners were also asked to listen to the recorded speech under the same condition when they judged the emotion. Since the noise was played through the headphones, we could record emotional speech acted in noise separately from the noise signals. Speech and noise were recorded onto two separate time-aligned tracks so that we can compute signal-to-noise ratio (SNR). The average SNR calculated from the separated tracks is around -8.7dB. 

\subsection{Experimental conditions}
\begin{figure}[t]
\centering
\includegraphics[width=8cm]{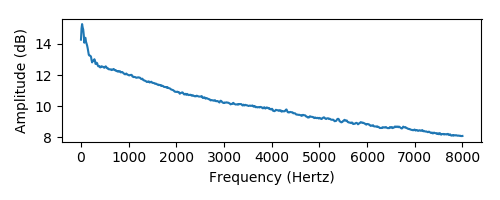}
\vspace{-0.8cm}
\caption{Long-term averaged spectrum of the noise}\label{fig:ltas}
\vspace{-0.5cm}
\end{figure}

The recording took place in a professional audio studio. Audio equipment included an AKG C314 condenser microphone and SHURE SE215 sound-isolating earphones. Three CueBoxes were used for audio monitoring by the listeners. Recordings were taken with a sampling frequency of 48 kHz and were 24-bit, and mono channel. Noise used for the recording and listener judgment was a mixture of speech-shaped noise called ICRA noise (6-person babble)~\cite{dreschler2001icra} and in-car noise. The long-term averaged spectrum of the noise is shown in Figure~\ref{fig:ltas}.


\section{Analysis on Acted and Perceived Emotions in Noise}
Since the database was recorded in different emotions and environmental conditions, 
it is important for us to understand how the emotions and environmental conditions affect each other.
First, a confusion matrix is employed to visualize the performance of acted and perceived emotions.
Second, the Frobenius distance between the confusion and identity matrices is calculated for further analysis.
Through utilizing the above measures, we aim to answer the following questions
\footnote{In addition, we carried out additional analysis to see whether the speakers made some changes or listeners learned some useful cues during the above recording procedure. Please see the appendix.}: a) Can speakers express their emotion correctly under both quiet and noisy conditions? b) Can listeners recognize the emotion contained in speech signals even in noise?

\subsection{Analysis based on confusion matrix}
\begin{figure}[t]
    \subfloat[Quiet]{
    \begin{minipage}{0.25\textwidth}
     \centering
     \includegraphics[width=4.3cm, trim={0cm 0cm 0cm 2cm}]{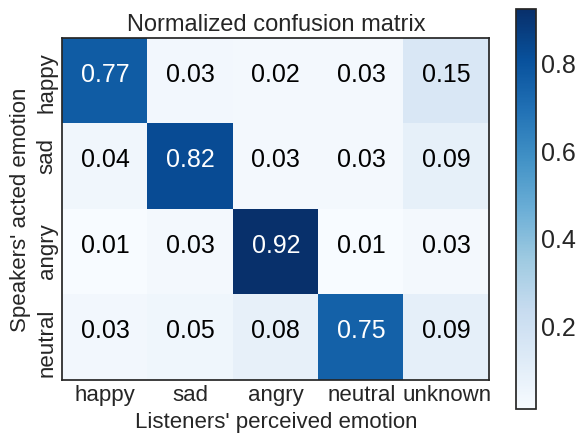}
   \end{minipage}}
    \subfloat[Noisy]{
    \begin {minipage}{0.25\textwidth}
     \centering
     \includegraphics[width=4.3cm, trim={0cm 0cm 0cm 2cm}]{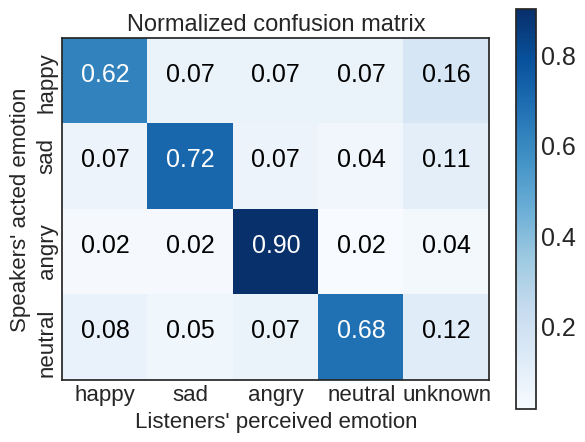}
   \end{minipage}}
   \vspace{-0.2cm}
   \caption{Confusion matrix of four types of emotional speech in quiet and noisy environments.} \label{fig:env_conf}
  
\end{figure}

\begin{figure}[t]
    \subfloat[High-level speakers (in quiet)]{
    \begin{minipage}{0.25\textwidth}
     \centering
     \includegraphics[width=4.3cm, trim={0cm 0cm 0cm 2cm}]{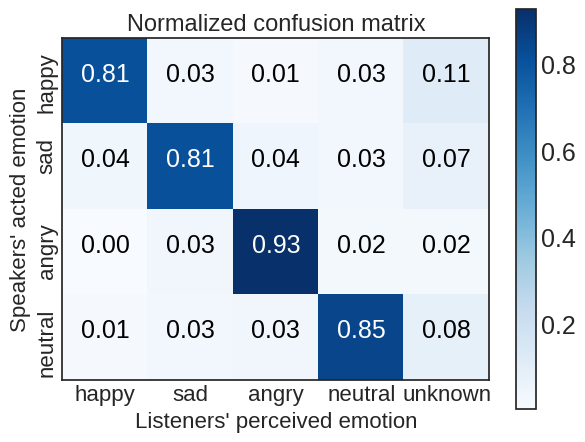}
  \end{minipage}}
    \subfloat[High-level speakers (in noisy)]{
        \begin {minipage}{0.25\textwidth}
     \centering
     \includegraphics[width=4.3cm, trim={0cm 0cm 0cm 2cm}]{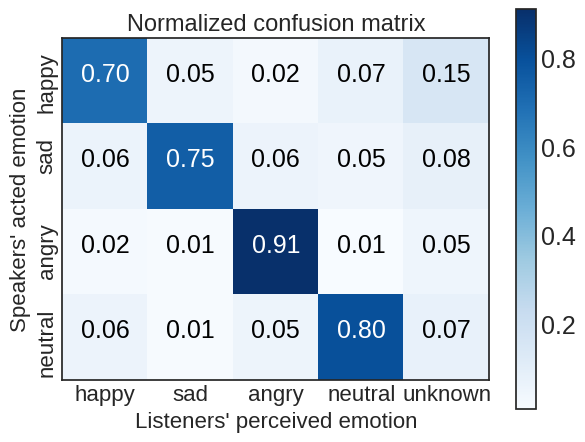}
  \end{minipage}}
  \\
  \subfloat[Low-level speakers (in quiet)]{
    \begin{minipage}{0.25\textwidth}
     \centering
     \includegraphics[width=4.3cm, trim={0cm 0cm 0cm 0.5cm}]{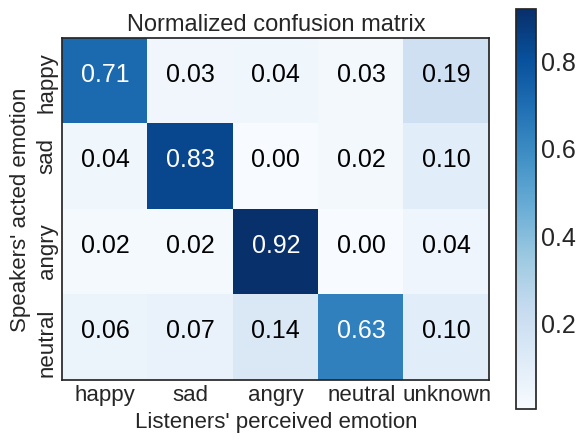}
  \end{minipage}}
    \subfloat[Low-level speakers (in noisy)]{
        \begin {minipage}{0.25\textwidth}
     \centering
     \includegraphics[width=4.3cm, trim={0cm 0cm 0cm 0.5cm}]{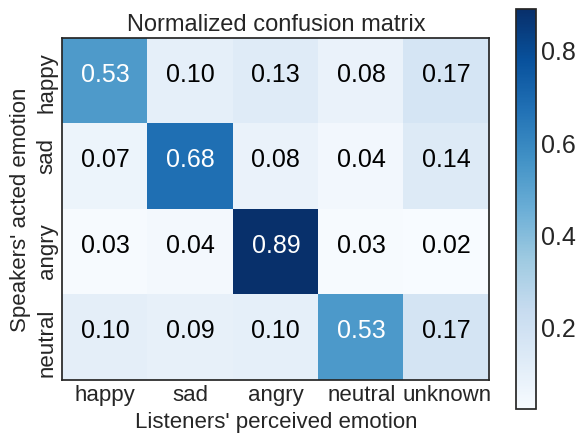}
  \end{minipage}}
  \vspace{-0.2cm}
  \caption{Confusion matrix of four types of emotional speech uttered by high-level and low-level speakers in quiet and noisy environments.} \label{fig:level_env_conf}
  \vspace{-0.5cm}
\end{figure}

Figure~\ref{fig:env_conf} shows the normalized confusion matrix in quiet and noisy environments, and Fig.~\ref{fig:level_env_conf} reveals the differences between high-level and low-level speakers under each environmental condition. In the confusion matrix, each row represents the acted emotions while each column stands for perceived emotions. From Fig.~\ref{fig:env_conf}, we find that the ratios of the correctly perceived emotional voices under the noise condition are obviously lower than those in the quiet condition. In the quiet environment, neutral tended to be the most confusing emotion, while in the noisy environment, the most confusing emotion was happy. Among the four emotions, anger was the least likely to be confusing in both quiet and noisy conditions and was least affected by environmental conditions. In contrast, happy voices were affected most by the environment. Figure~\ref{fig:level_env_conf} shows that high-level speakers could express the four types of emotional speech better than low-level speakers in both environments. We can see that the emotional speech uttered by high-level speakers resulted in much less confusion than that of low-level speakers, especially in the noisy environment. 

These results demonstrate that speakers can express their emotions correctly in both quiet and noisy environments. If speakers are better trained, they can produce more appropriate emotions robust to noisy conditions.

\subsection{Analysis based on Frobenius distance}
Frobenius distance is defined as the sum of the element-wise square differences of the abstraction of two matrices~\cite{laurent2012forecasting}. In order to compare the average accuracy of all emotions in different conditions, we calculate the Frobenius distance using 
\begin{equation}\label{eq:fd}
F = \mathbf{Tr}[(\mathcal{C}-\mathcal{I})^\top(\mathcal{C}-\mathcal{I})] 
\hfilneg 
\end{equation}
where $F$ is the Frobenius distance, $\mathcal{C}$ denotes the confusion matrix (except the `unknown` column), and $\mathcal{I}$ is a $4\times4$ identity matrix. The smaller the confusion of emotional speech, the smaller the distance. 

Table~\ref{tab:fro_spk_env_gen} shows the Frobenius distances computed separately per gender of speakers. Specifically, we computed a confusion matrix for each gender and computed its Frobenius distance to the identity matrix. From Table~\ref{tab:fro_spk_env_gen}, we can clearly see that gender is an important factor---emotional speech spoken by female speakers has a smaller Frobenius distance. In other words, it is less confusing than that uttered by male speakers.

However, this is not the case from the listener's perspective. According to Table~\ref{tab:fro_lis_env_gen}, the average distance computed separately per listener's gender is almost the same. On the other hand, it turns out that the listeners' age has a significant effect on the accuracy of their judgment, as we can see from Table~\ref{tab:fro_lis_age}. Listeners in their 20s and 30s were less confused than listeners in their 40s or 50s.

Also, we found that the performance is jointly influenced by the gender of speakers and listeners, as shown in Table~\ref{tab:fro_lis_age_gen}. Interestingly, the combination of male speakers and male listeners resulted in the worst performance of any combination.

As indicated by the above discussion, listeners were able to recognize the emotion contained in speech signals even under the noisy condition. However, the recognition accuracy largely depended on the listeners' age, and also had a relationship to the collocation of speakers’ and listeners’ gender.


\begin{table}[t]
\centering
\caption{Frobenius distance for \textbf{female and male speakers} under different conditions.}
\vspace{-0.1cm}
\label{tab:fro_spk_env_gen}
\begin{tabular}{c|ccc}
\toprule
\hline
Environment & Quiet & Noisy & Overall \\ \hline
Female      &   0.14    &  0.28     &    0.20  \\    \hline
Male        &    0.22   &     0.52  &     0.33 \\ \hline \bottomrule
\end{tabular}
\vspace{-0.1cm}
\end{table}

\begin{table}[!t]
\centering
\caption{Frobenius distance for \textbf{female and male listeners}.} \label{tab:fro_lis_env_gen}
\vspace{-0.1cm}
\begin{tabular}{c|ccc}
\toprule
\hline
Environment & Quiet & Noisy & Overall \\ \hline
Female      &   0.16    &  0.43     &    0.26  \\    \hline
Male        &    0.19   &     0.35  &     0.25 \\ \hline \bottomrule
\end{tabular}
\vspace{-0.1cm}
\end{table}

\begin{table}[!t]
\caption{Frobenius distance for listeners of different ages.} \label{tab:fro_lis_age}
\vspace{-0.1cm}
\centering
{%
\begin{tabular}{c|ccc}
\toprule
\hline
Environment & Quiet & Noisy & Overall \\ \hline
20s         & 0.10  & 0.30  & 0.17    \\ \hline
30s         & 0.10  & 0.25  & 0.16    \\ \hline
40s         & 0.23  & 0.55  & 0.35    \\ \hline
50s         & 0.42  & 0.64  & 0.50    \\ \hline \bottomrule
\end{tabular}%
}
\vspace{-0.1cm}
\end{table}


\begin{table}[!t]
\centering
\caption{Frobenius distance of \textbf{ the combinations of speaker's and listener's genders} (F = female, M = male).}\label{tab:fro_lis_age_gen}
\vspace{-0.1cm}
\begin{tabular}{c|cc|cc}
\toprule
\hline
Gender                              & \multicolumn{2}{c|}{Opposite} & \multicolumn{2}{c}{Same} \\ \hline
Speaker                             & F             & M             & F           & M           \\ \hline
Listener                            & M             & F             & F           & M           \\ \hline
\multirow{2}{*}{Frobenius distance} & 0.20          & 0.29          & 0.22        & 0.40        \\ \cline{2-5} 
                                    & \multicolumn{2}{c|}{0.23}     & \multicolumn{2}{c}{0.30} \\ \hline \bottomrule
\end{tabular}
\vspace{-0.4cm}
\end{table}


\section{Statistical Analysis of Acoustic Features}

\begin{figure}[t]
\vspace{-1.5mm}
    \subfloat[Mean value of F0]{
    \begin{minipage}{0.25\textwidth}
     \centering
     \includegraphics[scale=0.28,trim={3cm 0.2cm 2cm 1cm}]{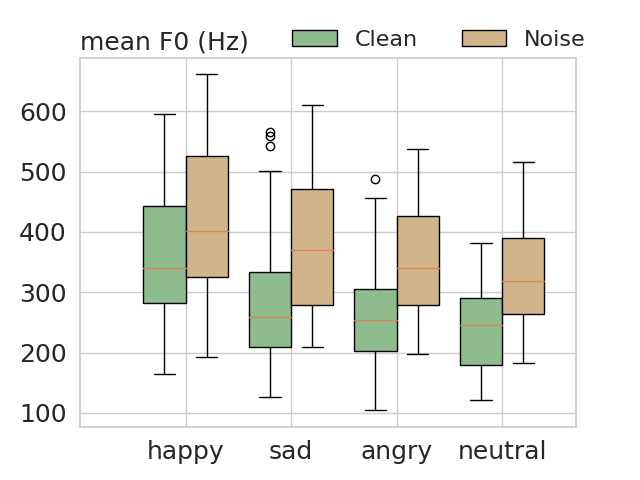}
  \end{minipage}}
    \subfloat[Spectral tilt]{
    \begin {minipage}{0.25\textwidth}
     \centering
     \includegraphics[scale=0.28,trim={3cm 0.2cm 2cm 1cm}]{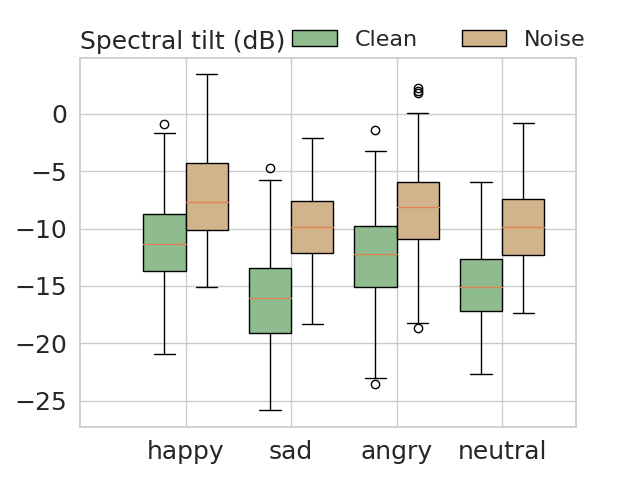}
  \end{minipage}}
  \\
    \subfloat[Mean value of HNR]{
    \begin {minipage}{0.25\textwidth}
     \centering
     \includegraphics[scale=0.28,trim={3cm 0.2cm 2cm 1cm}]{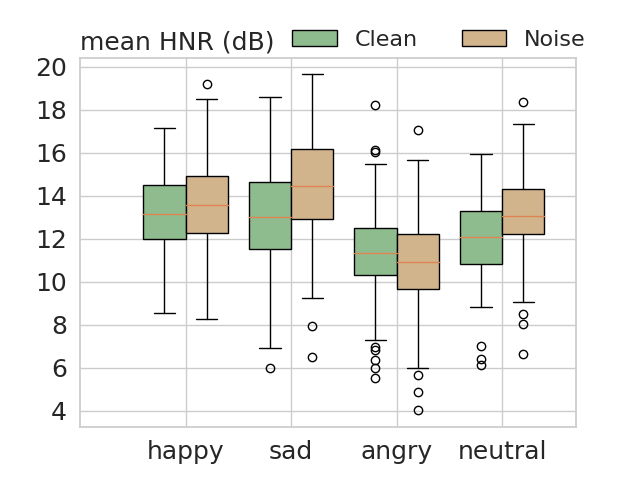}
  \end{minipage}}
  \subfloat[Mean value of intensity]{
    \begin{minipage}{0.25\textwidth}
     \centering
     \includegraphics[scale=0.28,trim={3cm 0.2cm 2cm 1cm}]{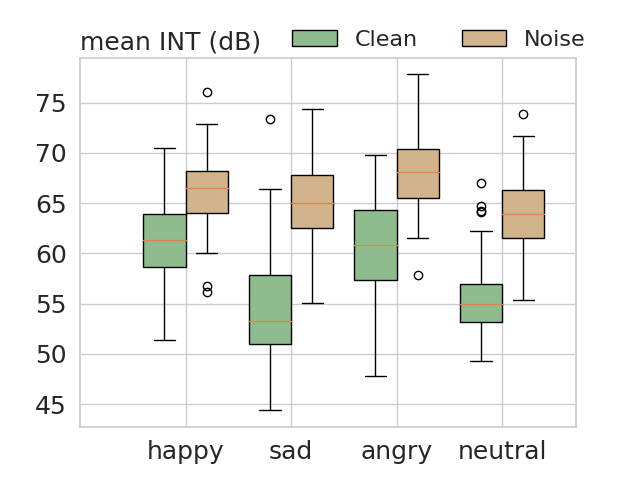}
  \end{minipage}}
  \\
    \subfloat[Mean value of F1]{
    \begin {minipage}{0.25\textwidth}
     \centering
     \includegraphics[scale=0.28,trim={3cm 0.5cm 2cm 1cm}]{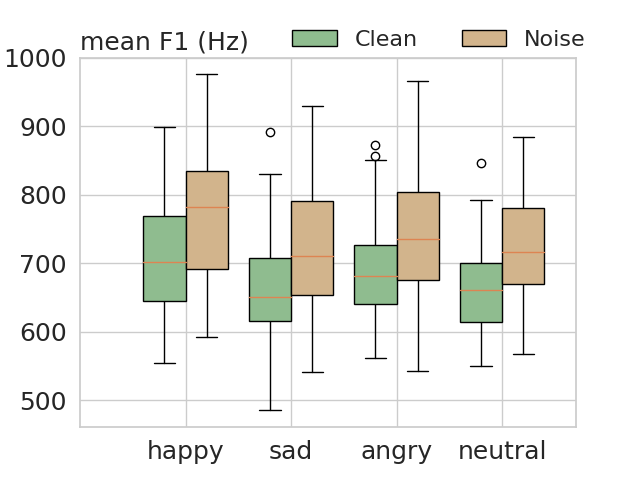}
  \end{minipage}}
  \subfloat[Mean value of F2]{
    \begin{minipage}{0.25\textwidth}
     \centering
     \includegraphics[scale=0.28,trim={3cm 0.5cm 2cm 1cm}]{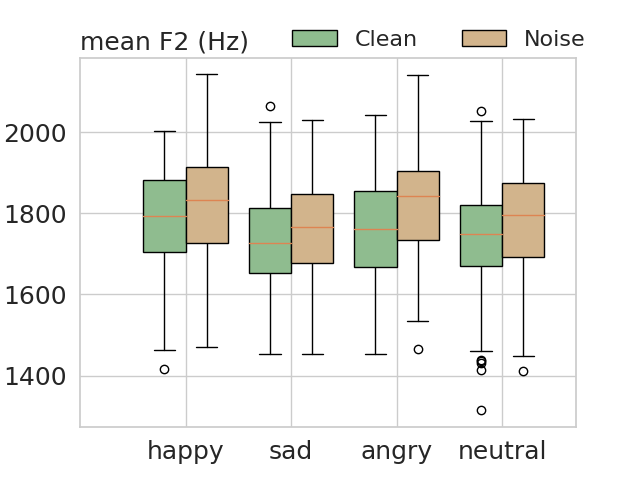}
  \end{minipage}}

  \caption{Boxplots of various acoustic features for each emotion in different environments.} \label{fig:acoustic_fea}
  \vspace{-0.6cm}
\end{figure}

Next, we analyzed the acoustic features of recorded emotional speech in quiet and noisy conditions to clarify which acoustic features are affected by noise. We used Parselmouth~\cite{jadoul2018introducing} to extract various acoustic features from each utterance, including F0, sound intensity~\cite{fahy1990sound}, harmonics-to-noise ratio (HNR)~\cite{boersma1993accurate}, the first and second formants (F1 and F2), and spectral tilt~\cite{lu2009contribution}. The distribution of each feature for each emotional category in both the quiet and noisy conditions is plotted in Fig.~\ref{fig:acoustic_fea}.

From these figures, we can see that our emotional speech data has typical patterns of the Lombard effect, such as higher F0 values, larger intensity, and flat spectral tilt. It is clear that even negative emotions (sad and angry) have higher pitch, interestingly. We can also see that since the Lombard effect increases F0 and intensity and makes spectral tilt flat in noise, relative differences of mean F0, intensity, and spectral tilt among emotional categories become smaller conversely. This is likely one reason why emotional speech in noise is more confusable. 

We conducted one-way ANOVA testing to see whether there are any statistically significant differences between the means of each feature of each emotion in quiet and noisy environments. The results suggest that if we set the significance level to 0.05, all features above are significantly different. 

We also conducted two-way ANOVA testing to see if there are further interactions between emotional categories and environmental conditions. We confirmed that emotional and environmental conditions are not independent factors and there are interactions regarding all the above features except F1 and F2 at the significance level of 0.05. This indicates that although the emotional speech clearly shows the typical pattern of the Lombard effect, the actual acoustic differences between emotional speech in quiet and noisy conditions are not constant and depend on the emotion category.

\begin{figure}[t]
 \centering
\includegraphics[width = 8cm,trim={2cm 5cm 5cm 5.5cm}]{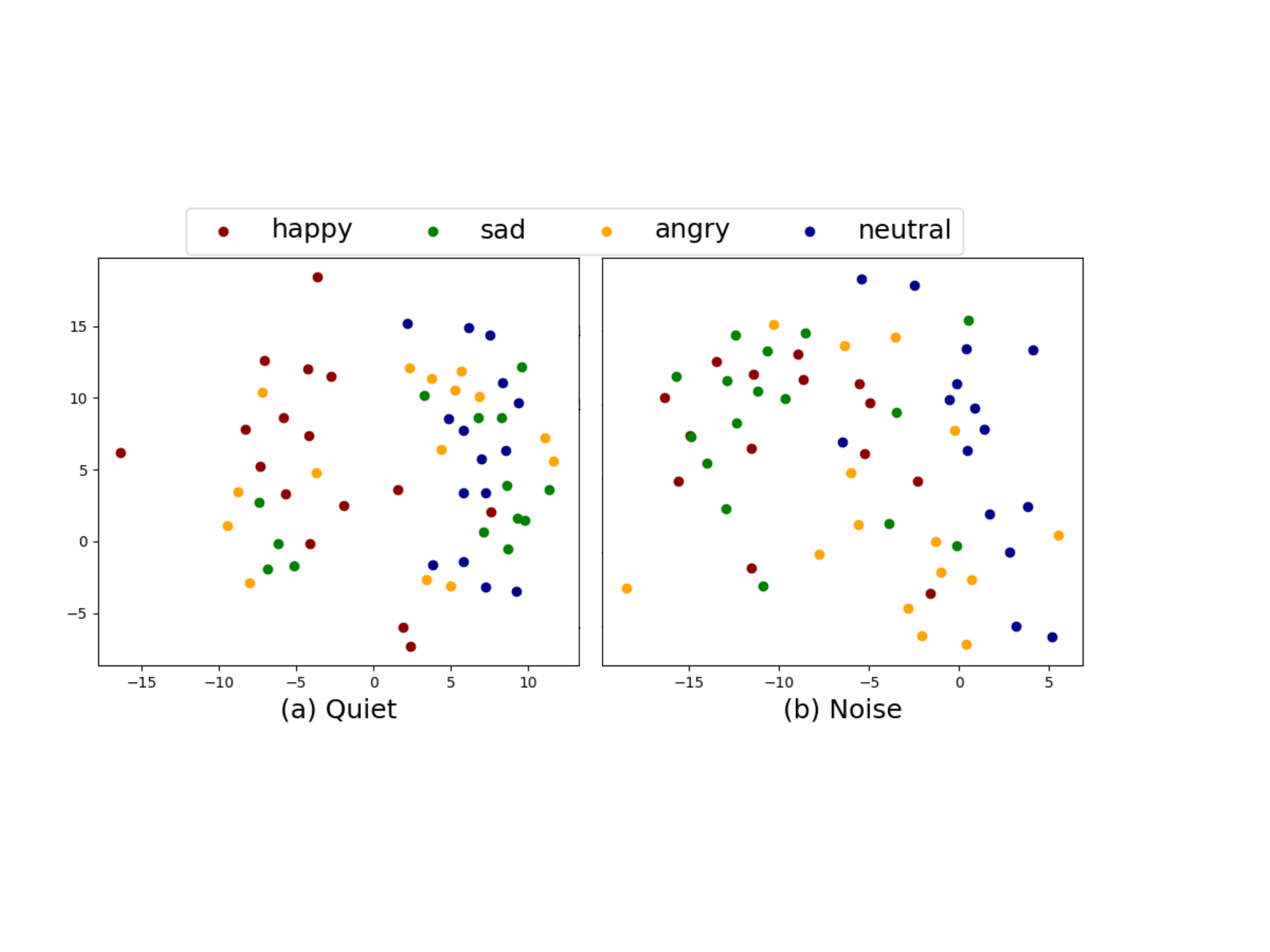}
\vspace{-0.8cm}
\caption{T-SNE visualization for acoustic features of one speaker's utterances. Each point represents one utterance.} \label{fig:2d_visual}
\vspace{-0.6cm}
\end{figure}

In addition to the mean value of each feature, we computed the maximum and minimum values as well as the variance of each feature for each utterance. We then put all these values into one array, and use t-SNE algorithm~\cite{maaten2008visualizing} to visualize them into 2D-dimensional space. The results of a high-level speaker are shown in Fig.\ \ref{fig:2d_visual}. By comparing the the quiet and noisy conditions shown in the figure, we can see that there are two separated clusters in the quiet condition. Happy and neutral speech samples are distributed in almost completely separate places. In the quiet condition, most points of sad speech are overlapped with many points of neutral speech. In contrast, in the noisy condition, many sad points are significantly overlapped with those of happy speech. 

\section{Speech Intelligibility}
Finally, in order to determine the influence of Lombard effect on speech intelligibility, we computed the short-time objective intelligibility (STOI)~\cite{taal2010short} scores, which is a standard objective intelligibility measure and its value ranges between 0 and 1. Since it clearly shows the tendency of the Lombard effect (as described earlier), we can hypothesize that emotional speech produced in noise would be more intelligible than that in clean. To verify the hypothesis, we first estimated the intelligibility of the recorded emotional speech without the Lombard effect. To calculate the STOI score, we added the same simulated noise discussed in Section 2 to emotional speech recorded in the quiet condition, and used their corresponding clean speech as the reference voice. We computed the STOI score of emotional speech acted in noise in the same way. As expected, the emotional speech produced in noise has a significantly higher STOI score (0.61) than the emotional speech produced in the quiet environment (0.44), which indicates that the emotional speech produced in noise is more intelligible.

\section{Summary}
To investigate the influence of the Lombard effect on emotional speech, we recorded parallel emotional utterances by 12 speakers under both quiet and noisy conditions in a professional recording studio. From the initial analysis results, we derive the following conclusions. 1) Speakers can express their emotion correctly even under adverse conditions. If speakers are better trained, they can produce more appropriate emotions robust to noisy conditions. Emotional speech uttered by female speakers is more accurately recognized than that of male speakers in general. 2) Younger listeners are able to recognize the emotion contained in speech signals under noise better than older listeners. 3) Emotional speech in noise shows the typical characteristics of the Lombard effect. However, the changes are complex: the acoustic differences between emotional speech in quiet and noisy environments depend on the emotion category. Finally, because of interactions with the Lombard effect, relative differences of important acoustic cues such as mean F0, intensity, and spectral tilt among emotion categories become smaller conversely. This is one reason why emotional speech in noise is more easily confused. 

As our next step, we will use the above findings on the joint effect of Lombard and emotions for speech emotion recognition and emotional speech synthesis.

\newpage
\bibliographystyle{IEEEtran}

\bibliography{main}
\newpage
\appendix
\section{Analysis of learning effect}
To see whether the speakers made some changes or listeners learned some useful cues during the above recording procedure, we plotted Fig.~\ref{fig:lis_utt}, in which the horizontal axis is the index of recorded utterances in chronological order, and the vertical axis is the number of listeners who correctly identified the acted emotion, divided by the total number of listeners. We fit their relationship with linear regression and calculated the correlation coefficient. In the quiet environment, the correlation coefficient of the angry speech was as high as 0.84, the sad speech was moderately correlated (0.55), and the neutral speech was weakly correlated (0.32). Meanwhile, in the noisy environment, the correlation coefficients of the four emotions were all different: 0.76 for neutral, 0.47 for happy, 0.45 for angry, and 0.15 for sad.

From this experiment, while we cannot conclude what was learned by the speakers and listeners, or why, we can at least tell that the identification ratios become obviously better through the listeners' feedback when it comes to angry and sad in quiet and neutral in noise. Interestingly, the identification ratios of happy speech in the quiet condition and sad speech in noise remained less correlated. The reasons for these phenomena require further investigation.

\begin{figure}[htbp]
\subfloat[Happy \textemdash{} Quiet]{
 \begin{minipage}{0.5\textwidth}
 \centering
 \includegraphics[scale=0.3,trim={6cm 0.5cm 5cm 1cm}]{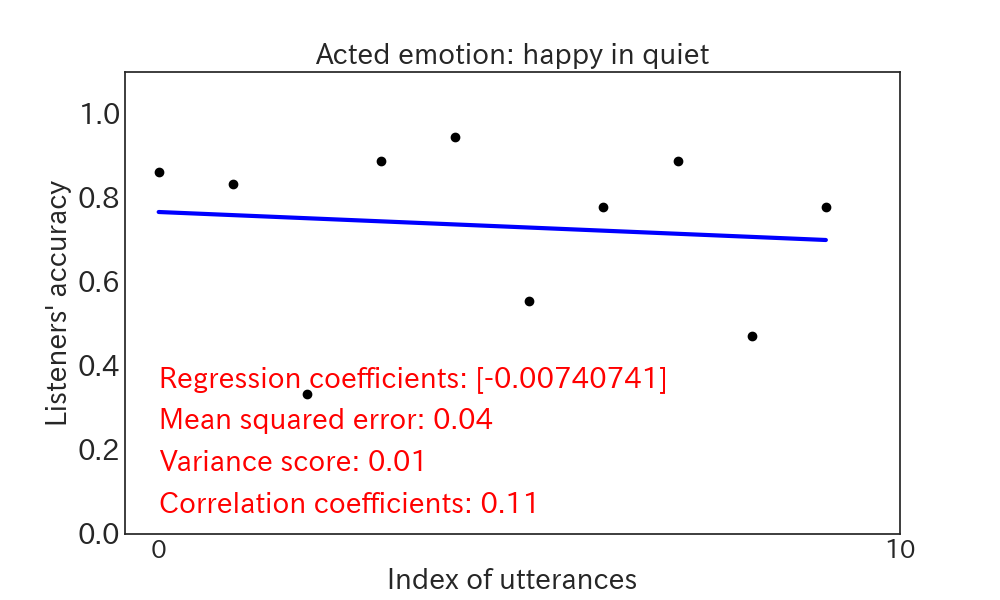}
 \end{minipage}}
 \\
\subfloat[Sad \textemdash{} Quiet]{
 \begin {minipage}{0.5\textwidth}
 \centering
 \includegraphics[scale=0.3,trim={6cm 0.5cm 5cm 1.2cm}]{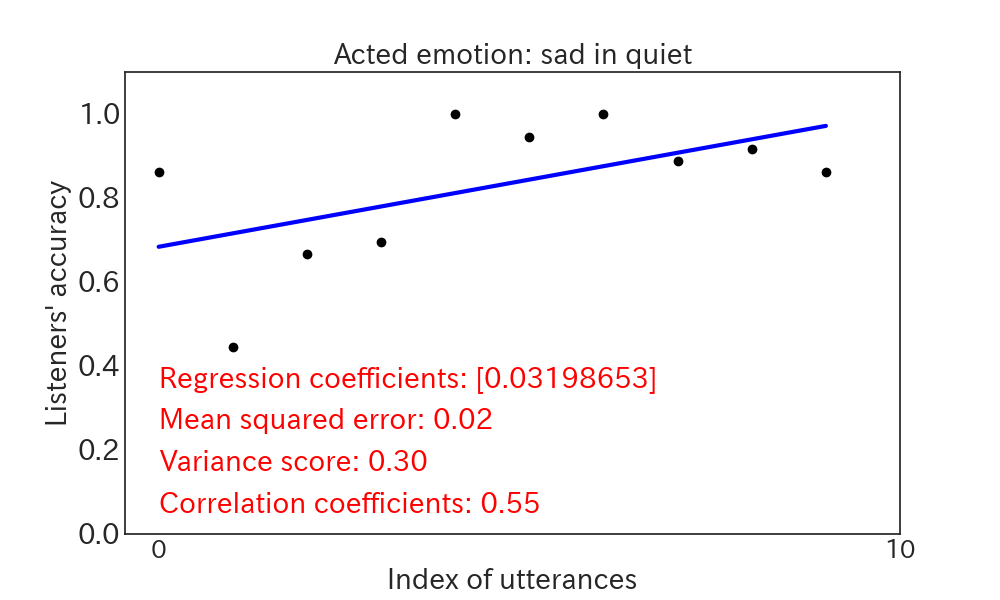}
 \end{minipage}}
 \\
\subfloat[Angry \textemdash{} Quiet]{
 \begin{minipage}{0.5\textwidth}
 \centering
 \includegraphics[scale=0.3,trim={6cm 0.5cm 5cm 1.2cm}]{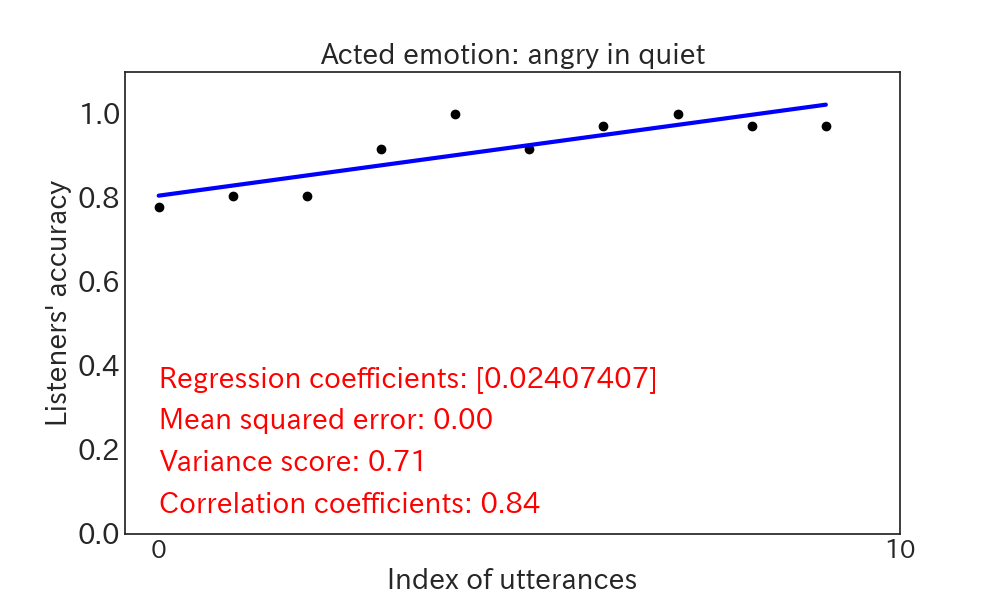}
 \end{minipage}}
\end{figure}

\begin{figure}[htbp]\ContinuedFloat
\subfloat[Neutral \textemdash{} Quiet]{
 \begin {minipage}{0.5\textwidth}
 \centering
 \includegraphics[scale=0.3,trim={6cm 0.5cm 5cm 1.2cm}]{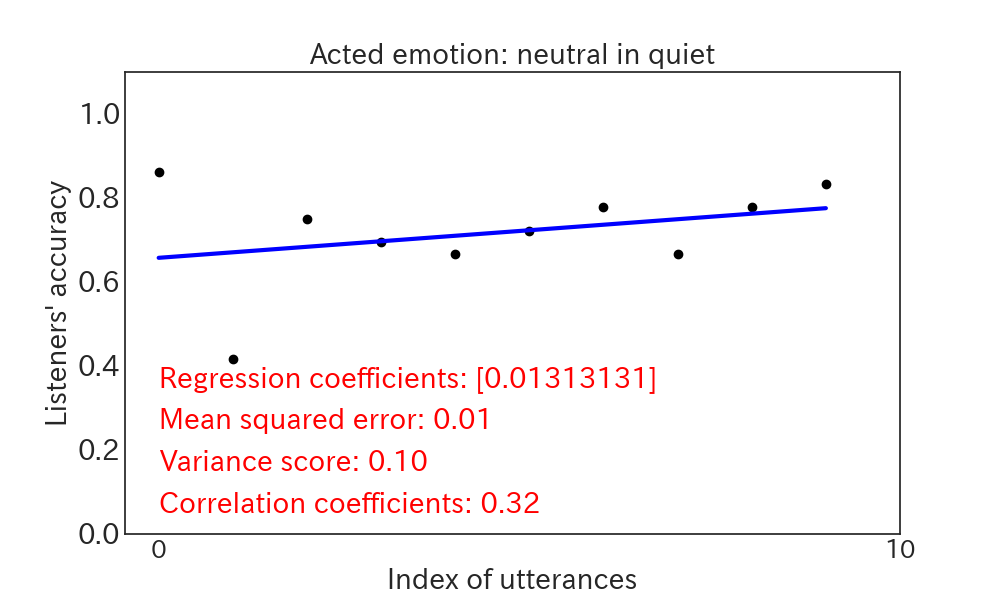}
 \end{minipage}}
 \\
\subfloat[Happy \textemdash{} Noise]{
 \begin{minipage}{0.5\textwidth}
 \centering
 \includegraphics[scale=0.3,trim={6cm 0.5cm 5cm 1.2cm}]{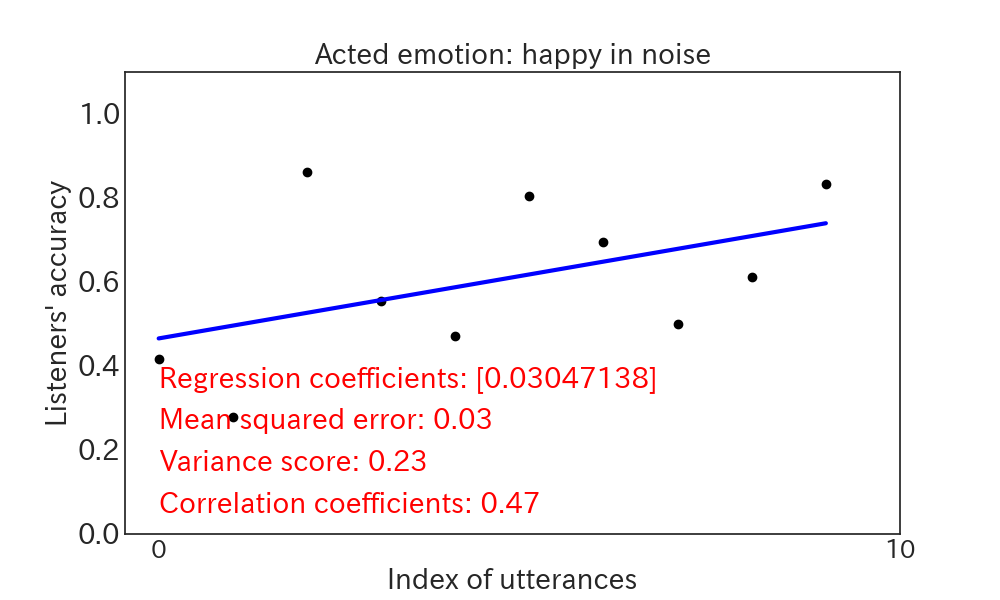}
 \end{minipage}}
 \hfill
\subfloat[Sad \textemdash{} Noise]{
 \begin {minipage}{0.5\textwidth}
 \centering
 \includegraphics[scale=0.3,trim={6cm 0.5cm 5cm 1.2cm}]{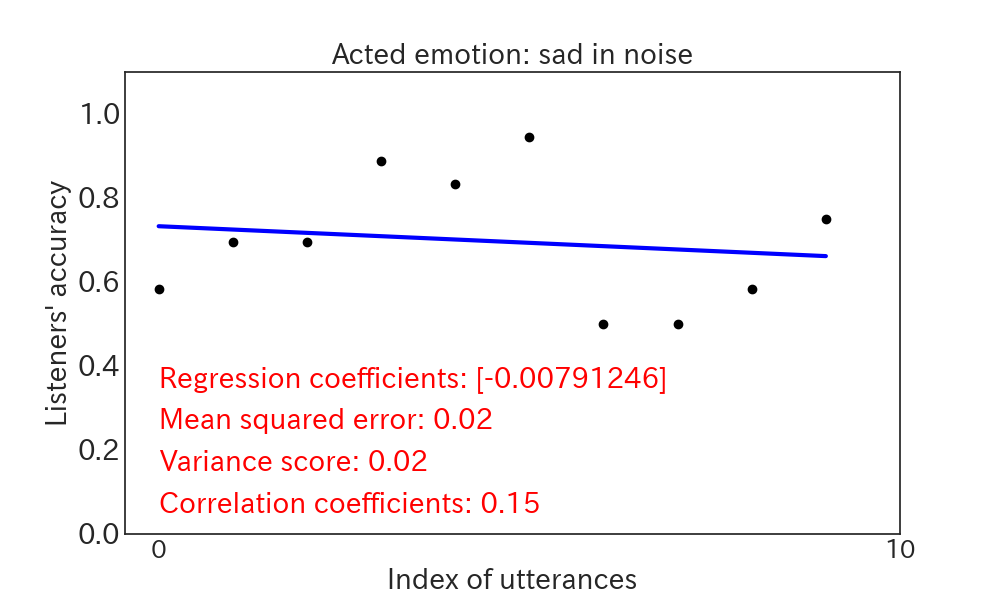}
 \end{minipage}}
 \\
\subfloat[Angry \textemdash{} Noise]{
 \begin{minipage}{0.5\textwidth}
 \centering
 \includegraphics[scale=0.3,trim={6cm 0.5cm 5cm 1.2cm}]{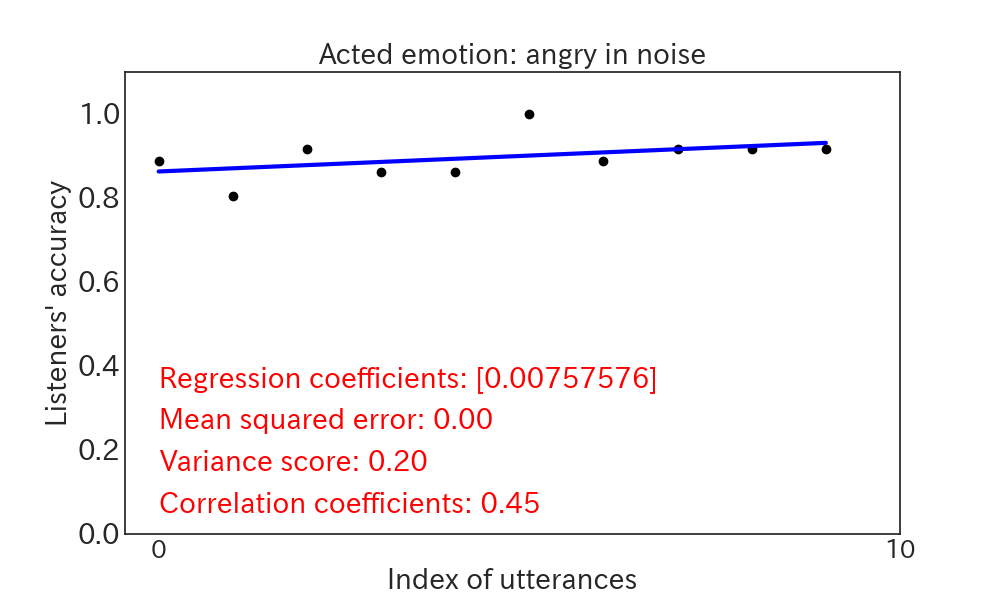}
 \end{minipage}}
 \hfill
\subfloat[Neutral \textemdash{} Noise]{
 \begin {minipage}{0.5\textwidth}
 \centering
 \includegraphics[scale=0.3,trim={5cm 0.5cm 5cm 1.2cm}]{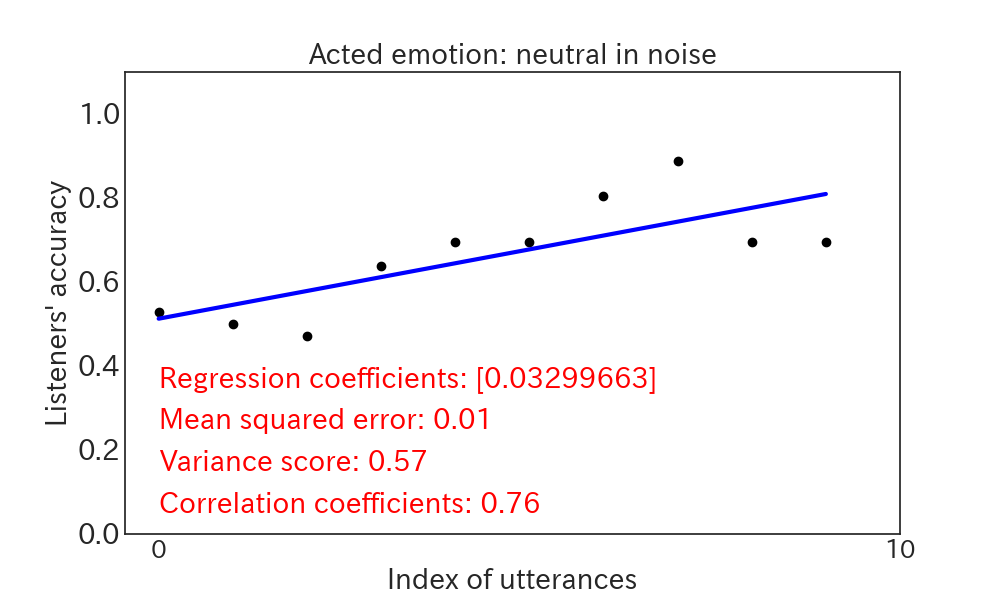}
 \end{minipage}} 
\caption{Linear regression for utterance index in chronological order and number of listeners who identified acted emotion correctly.} \label{fig:lis_utt}
\end{figure}
\end{document}